 \documentclass[final,authoryear,times,twocolumn]{elsarticle}

\usepackage{amssymb}

\biboptions{sort&compress}

\journal{Applied Radiation and Isotopes}

\begin{document}

\begin{frontmatter}



\title{Activation cross-sections of long lived products of deuteron induced nuclear reactions on dysprosium up to 50 MeV}


\author[1]{F. T\'ark\'anyi}
\author[1]{F. Ditr\'oi\corref{*}}
\author[1]{S. Tak\'acs}
\author[1]{J. Csikai}
\author[2]{A. Hermanne}
\author[3]{A.V. Ignatyuk}
\cortext[*]{Corresponding author: ditroi@atomki.hu}

\address[1]{Institute for Nuclear Research of the Hungarian Academy of Sciences (ATOMKI),  Debrecen, Hungary}
\address[2]{Cyclotron Laboratory, Vrije Universiteit Brussel (VUB), Brussels, Belgium}
\address[3]{Institute of Physics and Power Engineering (IPPE), Obninsk, Russia}

\begin{abstract}
Activation cross-sections for production of $^{162m,161,155}$Ho,$^{165,159,157,155}$Dy and $^{161,160,156,155}$Tb radionuclides in deuteron induced nuclear reactions on elemental dysprosium were measured up to 50 MeV for practical application and the test of the predictive power of nuclear reaction model codes. A stacked-foil irradiation technique and off-line $\gamma$-ray spectrometry were used to determine the activities. No earlier cross-section data were found in the literature. The experimental data are compared with the predictions of the ALICE-D, EMPIRE-D and TALYS codes. Integral production yields were calculated from the fitted experimental data.
\end{abstract}


\end{frontmatter}


\section{Introduction}
\label{1}
In the frame of our systematic study on the excitation functions of light charged particle induced nuclear reactions, the excitation functions for radionuclides production by deuteron irradiation of natDy were measured. Dysprosium is being a high priority and critical strategic metal now used world-wide for high technology, nuclear, clean energy and military applications. The knowledge on the activation cross-sections for light charged particles (ions) and neutron induced reactions are important for various applications.
The elemental dysprosium and some of its radionuclides are known to have applications in fission reactor physics and in medical field. 
The elemental dysprosium has a high thermal neutron absorption cross-section, which makes it interesting  for making neutron-absorbing control rods in nuclear reactors \citep{Kannan}, and a resonance absorber to control the coolant void reactivity in Advanced Heavy Water Reactor (AHWR) \citep{Kannan}. In medical treating damaged joints, irradiation with $^{165}$Dy has proved to be more effective than traditional surgery \citep{Srivastava}, and $^{159}$Dy has been advocated for transmission imaging and bone mineral analysis \citep{Rao}.  An alternative production route for radionuclides of lanthanides using high intensity lasers or light sources was studied recently by a group in Grenoble \citep{Habs}. 
There were several aims of the study
\begin{itemize}
\item	The earlier mentioned global task: the systematic study of the deuteron induced activation data, taking into account that no earlier data were found in the literature.
\item	Investigation of different production routes of the $^{161}$Ho medical radioisotope
\item	Investigation of the predictive power of widely used theoretical codes, via systematic comparison between the experimental and theoretical data  in connection with  the FENDL-3 Project (IAEA).
\item	To get experimental data in a wide mass region for (d,p) reactions to prepare systematics to be used in upgrading of nuclear reactions models for codes allowing better description of deuteron induced reactions.
\end{itemize}
Part of the experimental data for production of $^{161}$Ho was published recently \citep{Tarkanyi2013}. Here we repeat the data related to this radionuclide in graphical form for completeness and for comparison with the theories.

\section{Experiment and data evaluation}
\label{2}
The general characteristics and procedures for irradiation, activity assessment and data evaluation (including estimation of uncertainties) were similar to those in our earlier works \citep{Takacs,Tarkanyi2012}. The used high purity target foils (\textless 98\% purity) were bought from Goodfellow\copyright\  and were handled according to the producer's recommendations. The foil  thicknesses and homogeneity were accurately re-measured.
The main experimental parameters for the present study are summarized in Table 1 \citep{Andersen, Bonardi, Canberra, Error, Kinsey, Pritychenko, Forgamma, Tarkanyi1991, Tarkanyi2001} together with the methods used in data evaluation. The used decay data and Q values of the contributing reactions are collected in Table 2. To illustrate the reliability of the incident energy and the intensity of the deuteron beam along the stack, the simultaneously measured complete excitation function of the monitor reaction $^{27}$Al(d,x)$^{24}$Na is shown in Fig. 1 in comparison with the recommended data taken from \citep{Tarkanyi2001}.

\begin{table*}[t]
\tiny
\caption{Main experimental parameters and methods of data evaluation}
\centering
\begin{center}
\begin{tabular}{|p{0.9in}|p{1.4in}|p{0.9in}|p{1.5in}|} \hline 
\textbf{Parameter} & \textbf{Value} & \textbf{Data evaluation steps} & \textbf{Used method, software, database} \\ \hline 
Incident particle & Deuteron  & Gamma spectra evaluation & Genie 2000 \citep{Canberra}, Forgamma \citep{Forgamma} \\ \hline 
Measurement method & Stacked foil & Determination of beam intensity & Faraday cup (preliminary)\newline Fitted monitor reaction (final)\citep{Tarkanyi1991} \\ \hline 
Stack composition & Tb-Al-Dy, repeated 15 times & Decay data & NUDAT 2.6 \citep{Kinsey}   \\ \hline 
Target and thickness  & ${}^{nat}$Dy foil, 100.59 $\mu$m & Reaction Q-values & Q-value calculator \citep{Pritychenko} \\ \hline 
Number of targetfoils & 15 (Dy) & Determination of  beam energy & Andersen (preliminary) \citep{Andersen}\newline Fitted monitor reaction (final) \citep{Tarkanyi1991} \\ \hline 
Accelerator & Cyclone 90 cyclotron of the UniversitéCatholique in Louvain la Neuve (LLN)  & Uncertainty of energy & cumulative effects of possible uncertainties \\ \hline 
Primary energy & 50 MeV & Cross-sections & Elemental cross-section \\ \hline 
Irradiation time & 30 min & Uncertainty of cross-sections & sum in quadrature of all individual contributions \citep{Error} \\ \hline 
Beam current & 120 nA & Yield & Physical yield \citep{Bonardi} \\ \hline 
Monitor reaction, [recommended values]  & ${}^{27}$Al(d,x)${}^{24}$Na  reaction (\citep{Tarkanyi2001} &  &  \\ \hline 
Monitor target and thickness & ${}^{nat}$Al, 26.96 $\mu$m &  &  \\ \hline 
detector & HpGe &  &  \\ \hline 
g-spectra measurements & 4 series &  &  \\ \hline 
Cooling times & 4h, 30h, 430, 2380h &  &  \\ \hline 
\end{tabular}
\end{center}
\end{table*}

\begin{table*}[t]
\tiny
\caption{Decay data and contributing reactions}
\centering
\begin{center}
\begin{tabular}{|p{0.9in}|p{0.6in}|p{0.6in}|p{0.4in}|p{0.9in}|p{0.8in}|} \hline 
Nuclide & Half-life & E${}_{\gamma}$(keV) & I${}_{\gamma}$(\%) & Contributing reaction & Q-value\newline (keV) \\ \hline 
\textbf{${}^{16}$${}^{2m}$Ho\newline }IT: 62 \%\textbf{\newline } & 67.0 min & 57.74 & 4.4 & ${}^{161}$Dy(d,n)\newline ${}^{162}$Dy(d,2n)\newline ${}^{163}$Dy(d,3n)\newline ${}^{164}$Dy(d,4n) & 3050.38\newline -5146.61\newline -11417.62 \\ \hline 
\textbf{${}^{16}$${}^{1}$Ho\newline }$\varepsilon $: 100 \%\textbf{} & 2.48 h & 77.42\newline 103.05\newline 157.26\newline 175.42 & 1.9\newline 103.05\newline 0.49\newline 0.43 & ${}^{160}$Dy(d,n)\newline ${}^{161}$Dy(d,2n)\newline ${}^{162}$Dy(d,3n)\newline ${}^{163}$Dy(d,4n)\newline ${}^{164}$Dy(d,5n) & 2589.18\newline -3865.2\newline -12062.2\newline -18333.21\newline -25991.33 \\ \hline 
\textbf{$^{155}$Ho\newline }$\varepsilon + \beta^{+}$\textbf{} & 48~min~ & 103.89~\newline 136.30~\newline 240.19\newline 325.40~ & 2.17~\newline 5.00~\newline 12.5\newline 2.75~ & ${}^{158}$Dy(d,5n)\newline ${}^{160}$Dy(d,7n)\newline ${}^{161}$Dy(d,8n)\newline ${}^{162}$Dy(d,9n)\newline ${}^{163}$Dy(d,10n)\newline ${}^{164}$Dy(d,11n) & -31593.3\newline -47001.9\newline -53456.3\newline -61653.3\newline ? \\ \hline 
\textbf{${}^{165}$Dy\newline }$\beta $${}^{-}$: 100 \%\textbf{} & 2.334 h & 94.700\newline 279.763\newline 361.68\newline 633.415\newline 715.328 & 3.80\newline 0.534\newline 0.904\newline 0.613\newline 0.578 & ${}^{164}$Dy(d,p) & 3491.394 \\ \hline 
\textbf{${}^{159}$Dy\newline }$\varepsilon $: 100 \%\textbf{} & 144.4 d & 58.0 & 2.27 & ${}^{158}$Dy(d,p)\newline ${}^{160}$Dy(d,p2n)\newline ${}^{161}$Dy(d,p3n)\newline ${}^{162}$Dy(d,p4n)\newline ${}^{163}$Dy(d,p5n)\newline ${}^{164}$Dy(d,p6n)\newline ${}^{159}$Ho decay & 4608.11\newline -10800.48\newline -17254.87\newline -25451.86\newline -31722.87\newline -39380.99 \\ \hline 
\textbf{${}^{157}$Dy\newline }$\varepsilon $: 100 \%~\textbf{} & 8.14 h & 182.424\newline 326.336 & 1.33\newline 93 & ${}^{158}$Dy(d,p2n)\newline ${}^{160}$Dy(d,p4n)\newline ${}^{161}$Dy(d,p5n)\newline ${}^{162}$Dy(d,p6n)\newline ${}^{163}$Dy(d,p7n)\newline ${}^{164}$Dy(d,p8n)\newline ${}^{157}$Ho decay & -11280.12\newline -26688.71\newline -33143.1\newline -41340.09\newline -47611.11\newline -55269.23 \\ \hline 
\textbf{${}^{155}$Dy\newline }$\varepsilon $: 100 \%~\textbf{} & 9.9 h & 184.564\newline 226.918 & 3.37 \newline 68.4 & ${}^{158}$Dy(d,p4n)\newline ${}^{160}$Dy(d,p6n)\newline ${}^{161}$Dy(d,p7n)\newline ${}^{162}$Dy(d,p8n)\newline ${}^{163}$Dy(d,p9n)\newline ${}^{164}$Dy(d,p10n)\newline ${}^{155}$Ho decay & -27690.7\newline -43099.3\newline -49553.7\newline -57750.7\newline -64021.7 \\ \hline 
\textbf{${}^{161}$Tb\newline }$\beta $${}^{-}$: 100 \%~\textbf{} & 6.89 d & 74.56669\newline 87.941\newline 103.065\newline 106.113\newline 292.401 & 10.2\newline 0.183\newline 0.101\newline 0.078\newline 0.058 & ${}^{161}$Dy(d,2p)\newline ${}^{162}$Dy(d,2pn)\newline ${}^{163}$Dy(d,2p2n)\newline ${}^{164}$Dy(d,2p3n)\newline  & -2035.16\newline -10232.16\newline -16503.17\newline -24161.29 \\ \hline 
\textbf{${}^{160}$Tb\newline }$\beta $${}^{-}$:100 \%~\textbf{} & 72.3 d & 86.7877\newline 298.5783\newline 879.378\newline 966.166\newline 1177.954 & 13.2\newline 26.1\newline 30.1\newline 25.1\newline 14.9 & ${}^{160}$Dy(d,2p)\newline ${}^{161}$Dy(d,2pn)\newline ${}^{162}$Dy(d,2p2n)\newline ${}^{163}$Dy(d,2p3n)\newline ${}^{164}$Dy(d,2p4n) & -3277.34\newline -9731.73\newline -17928.73\newline -24199.74\newline -31857.86 \\ \hline 
\textbf{${}^{156}$Tb\newline }$\varepsilon $: 100 \%~\textbf{} & 5.35 d & 88.97\newline 199.19\newline 262.54\newline 296.49\newline 356.38\newline 422.34\newline 534.29\newline 1065.11\newline 1154.07\newline 1222.44 & 18\newline 41\newline 5.8\newline 4.5\newline 13.6\newline 8.0\newline 67\newline 10.8\newline 10.4\newline 31 & ${}^{158}$Dy(d,2p2n)\newline ${}^{160}$Dy(d,2p4n)\newline ${}^{161}$Dy(d,2p5n)\newline ${}^{162}$Dy(d,2p6n)\newline ${}^{163}$Dy(d,2p7n)\newline ${}^{164}$Dy(d,2p8n) & -17899.45\newline -33308.04\newline -39762.43\newline -47959.42\newline -54230.44\newline -61888.55 \\ \hline 
\textbf{${}^{155}$Tb\newline }$\varepsilon $: 100 \%\textbf{} & 5.32 d & 86.55\newline 105.318\newline 148.64\newline 161.29\newline 163.28\newline 180.08\newline 262.27 & 32.0\newline 25.1\newline 2.65\newline 2.76\newline 4.44\newline 7.5\newline 5.3 & ${}^{158}$Dy(d,2p3n)\newline ${}^{160}$Dy(d,2p5n)\newline ${}^{161}$Dy(d,2p6n)\newline ${}^{162}$Dy(d,2p7n)\newline ${}^{163}$Dy(d,2p8n)\newline ${}^{155}$Dy decay & -24813.9\newline -40222.5\newline -46676.9\newline -54873.9\newline -61144.9 \\ \hline 
\end{tabular}
\end{center}
\begin{flushleft}
\tiny
\noindent Naturally occurring dysprosium is composed of 7 isotopes (${}^{156}$Dy -0.06 \%, ${}^{158}$Dy-0.10 \%, ${}^{160}$Dy-2.34 \%, ${}^{161}$Dy-18.9 \%, ${}^{162}$Dy-25.5 \%, ${}^{163}$Dy-24.9 \% and ${}^{164}$Dy-28.2 \%).
\noindent When complex particles are emitted instead of individual protons and neutrons the Q-values have to be decreased by the respective binding energies of the compound particles: np-d, +2.2 MeV; 2np-t, +8.48 MeV; n2p-${}^{3}$He, +7.72 MeV; 2n2p-a, +28.30 MeV
\end{flushleft}
\end{table*}

\begin{figure}[h]
\includegraphics[scale=0.3]{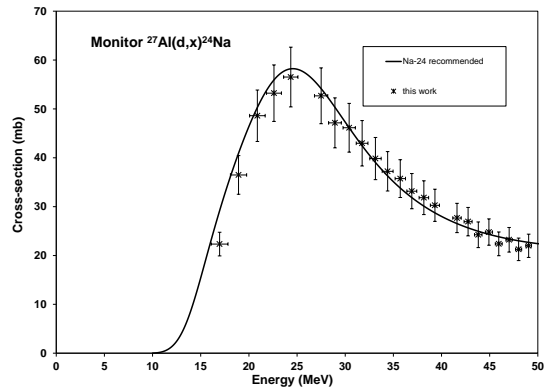}
\caption{The simultaneously measured monitor reactions for determination of deuteron beam energy and intensity}
\end{figure}

\section{Theory}
\label{3}
The ALICE-IPPE-D and EMPIRE-D \citep{Dityuk, Herman} calculations were performed using the recommended values for the input parameters. The results of TALYS \citep{Koning2007}  code were taken from the TENDL 2012 library \citep{Koning2012}.
The ALICE-D and EMPIRE-D are the modified version of ALICE-IPPE  \citep{Dityuk} and EMPIRE \citep{Herman} codes and were developed at IPPE (Institute of Physics and Power Engineering, Obninsk, Russia) for better description of activation cross-sections of deuteron induced reactions by using energy dependent enhancement factor \citep{Ignatyuk} for simulation of direct (d,p) and (d,t) transitions. Independent data for isomers with ALICE-D code was obtained by using the isomeric ratios calculated with EMPIRE-D.
Separate sets of calculations were performed for all stable Dy isotopes as target nuclei. The excitation functions for natural isotopic composition were obtained by summing all these individual excitation functions with weight of their respective natural abundances.

\section{Results}
\label{4}

\subsection{Cross-sections}
\label{4.1}
Activation cross-sections for production of the $^{162m,161,155Ho,165,159,157,155}$Dy and $^{161,160,156,155}$Tb radionuclides were measured. The experimental cross-section data are shown in Figs. 2-11 in comparison with the predictions of theoretical codes. The numerical data are collected in Tables 3 and 4. The cross-section values of holmium production are for direct production via (d,xn) reactions. The dysprosium radio-isotopes are produced directly via (d,pxn) reactions or additionally through the decay of the shorter-lived isobaric parent holmium radioisotope (cumulative cross-section). The terbium radioisotopes are produced directly via (d,2pxn) reaction (including complex particle emission), or additionally from the decay of parent dysprosium or  holmium radioisotopes. Taking into account that the contribution from the decay of the holmium radioisotopes is low, these contributions were neglected in the presented theoretical results. The ground-state of the produced radioisotopes out of the direct production can be produced additionally through the internal transition of the isomeric state. The cross-section is marked with (m+), when the half-life of the isomeric state is significantly shorter compared to the half-life of the ground state and the cross-section of the  production of ground state were deduced from spectra after nearly complete decay of the isomeric state.  

\subsubsection{Cross-sections for the $^{nat}$Dy(d,x)$^{162m}$Ho reaction}
\label{4.1.1}
The production cross-sections of $^{162m}$Ho (T$_{1/2}$=67.0 min) are shown in Fig. 2. None of the theoretical calculations represents well the experimental data and large disagreements exist among the codes. TENDL-2012 values, which are the results of most recent version of TALYS are a factor of 2 higher and shifted to higher energies compared to the previous results.

\begin{figure}[h]
\includegraphics[scale=0.3]{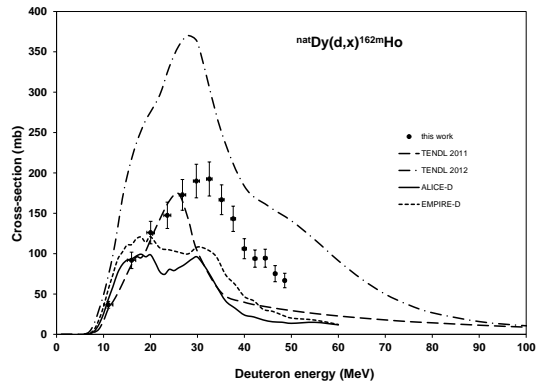}
\caption{Experimental and theoretical excitation functions for $^{nat}$Dy(d,x)$^{162m}$Ho}
\end{figure}

\subsubsection{Cross-sections for the $^{nat}$Dy(d,x)$^{161g}$Ho(m+) reaction}
\label{4.1.2}
The measured cross-section include the full contribution of the decay of the short-lived isomer (T$_{1/2}$ = 6.7 s). According to Fig. 3, the shape of the experimental and theoretical exaction functions of $^{161}$Ho (T$_{1/2}$= 2.48 h) are similar, but the overestimation of all codes is surprisingly high. For the TENDL results (both libraries give about the same values) a scaling factor of 0.55 results in a reasonable representation of the experiment.

\begin{figure}[h]
\includegraphics[scale=0.3]{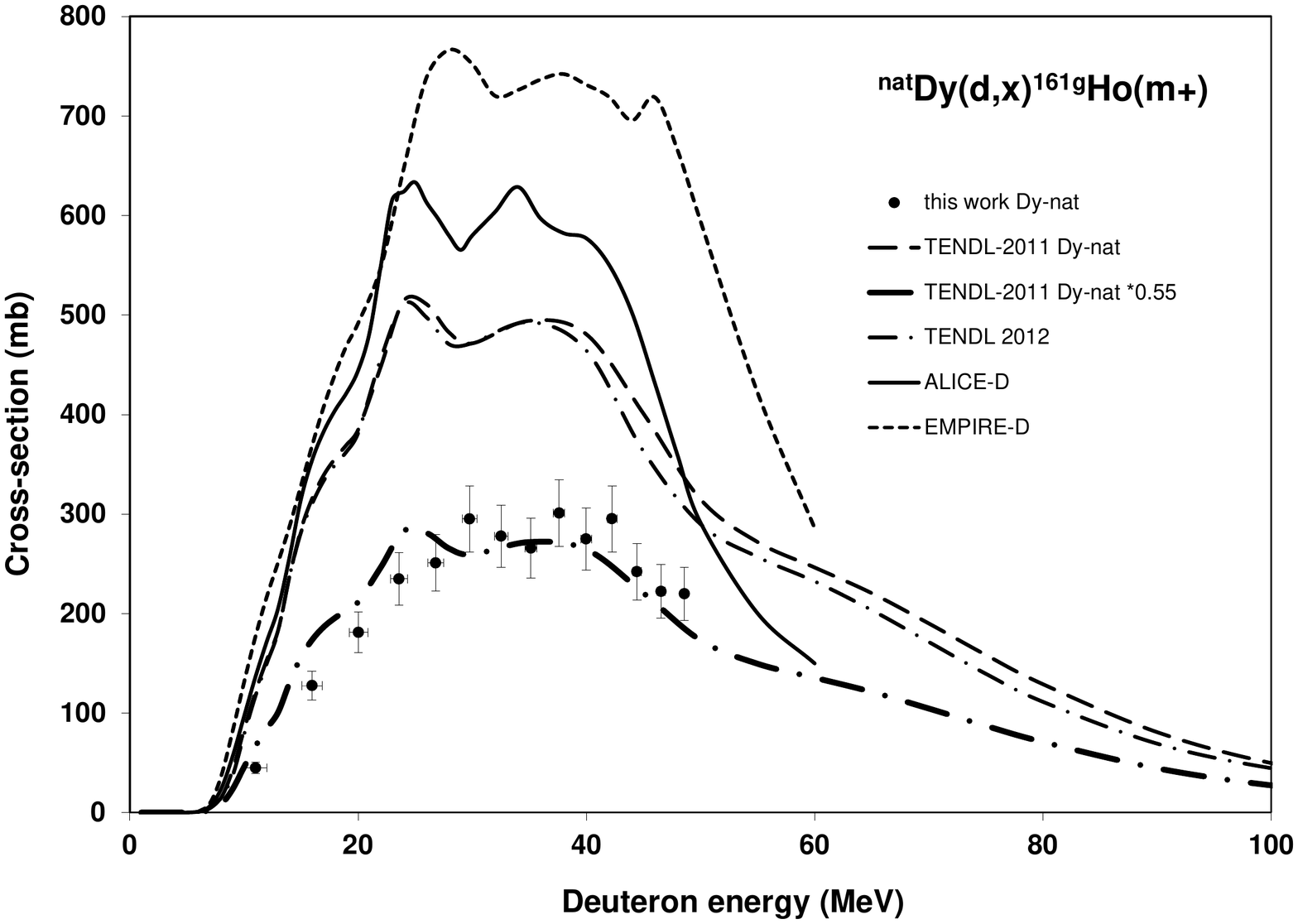}
\caption{Experimental and theoretical excitation functions for $^{nat}$Dy(d,x)$^{161g}$Ho(m+)}
\end{figure}

\subsubsection{Cross-sections for the $^{nat}$Dy(d,x)$^{165g}$Dy(m+) reaction}
\label{4.1.3}
$^{165}$Dy (T$_{1/2}$= 2.334 h) is produced only via the $^{164}$Dy(d,p) reaction. The measured cross-section contains the contribution of the complete decay of the shorter-lived isomeric state (T$_{1/2}$=1.257 min) that decays with IT (97.74 \%) to the ground state (m+). 
The experimental data are in good agreement with the systematics of the experimental (d,p) cross-sections above 40 MeV and between the predictions of EMPIRE-D and ALICE-D. The new TENDL 2012 data are closer to the experiment than TENDL 2011 but the underestimation is still significant.

\begin{figure}[h]
\includegraphics[scale=0.3]{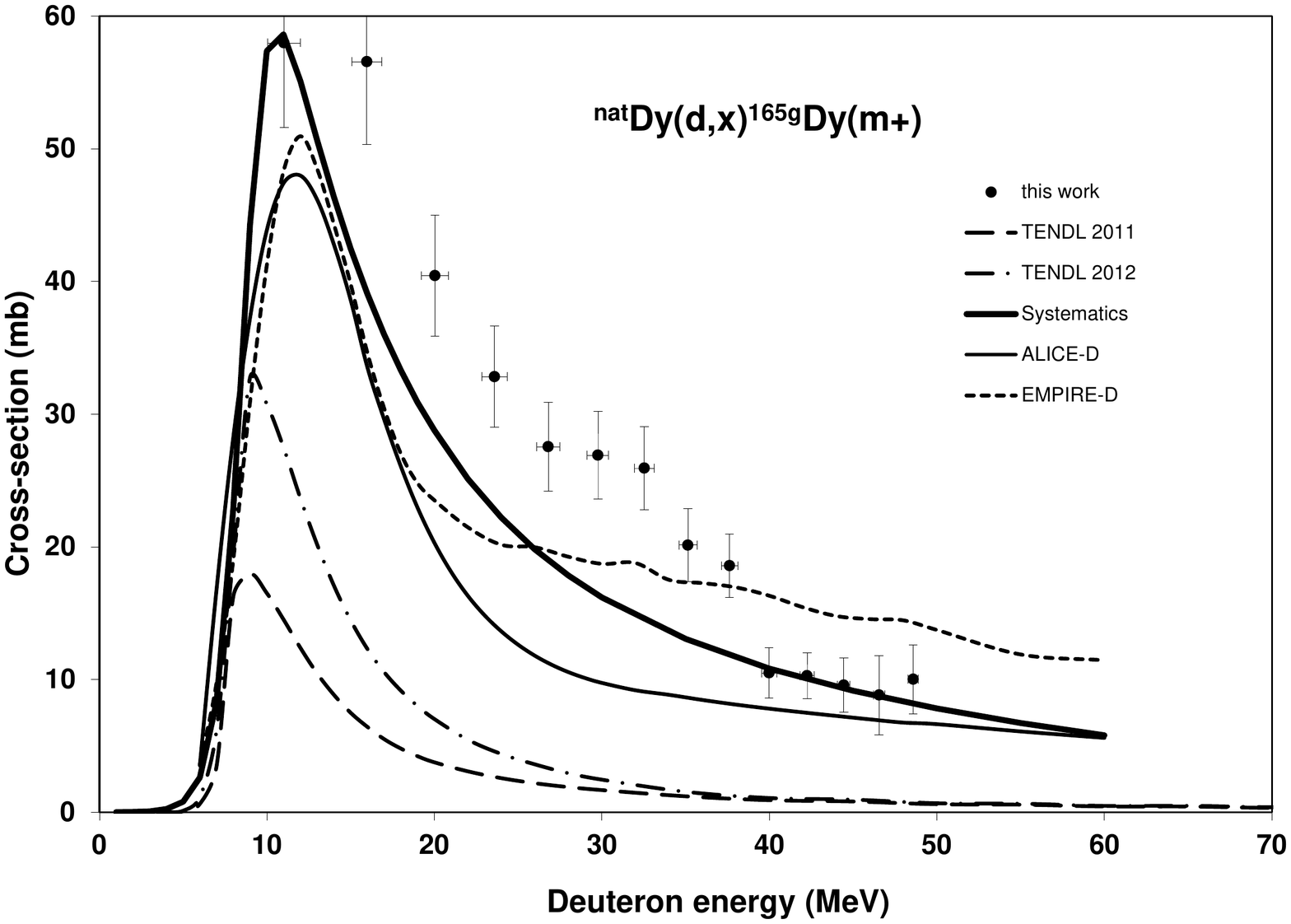}
\caption{Experimental and theoretical excitation functions for $^{nat}$Dy(d,x)$^{165g}$Dy(m+)}
\end{figure}

\subsubsection{Cross-sections for the $^{nat}$Dy(d,x)$^{159}$Dy(cum) reaction}
\label{4.1.4}
The cumulative cross-sections of $^{159}$Dy(T$_{1/2}$ = 144.4 d) include the direct production and the contribution from the decay of $^{159}$Ho (T$_{1/2}$ = 33.05 min), were impossible to measure in our experimental conditions (too long cooling time). The overestimations of the cumulative cross-sections given in both TENDL libraries and by the ALICE-D and EMPIRE-D codes are significant (Fig. 5).

\begin{figure}[h]
\includegraphics[scale=0.3]{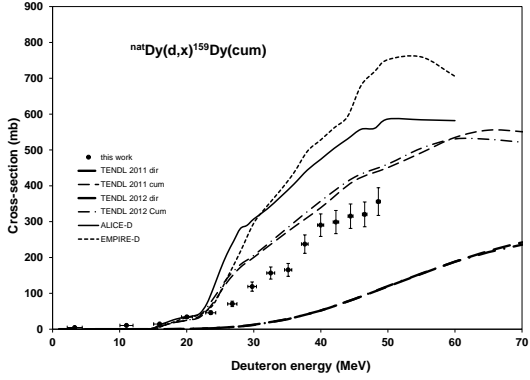}
\caption{Experimental and theoretical excitation functions for $^{nat}$Dy(d,x)$^{159}$Dy(cum)}
\end{figure}

\subsubsection{Cross-sections for the $^{nat}$Dy(d,x)$^{157}$Dy (cum) reaction}
\label{4.1.5}
Under our experimental circumstances we can deduce only cumulative cross-sections for production of $^{157}$Dy (T$_{1/2}$ = 8.14 h) including the contribution of total decay of short-lived parent $^{157}$Ho (T$_{1/2}$ = 12.6 min). The three codes reproduce well the complex shape of the cumulative experimental excitation function (direct on multiple stable target isotopes at higher energy and parent decay below 15 MeV) (see Fig. 6), but are one order of magnitude lower for the $^{156}$Dy(d,n)$^{157}$Ho reaction. At higher energies all codes overestimate the experimental results by a factor of two. According to the theory the direct reaction contributes only by about 20\%.

\begin{figure}[h]
\includegraphics[scale=0.3]{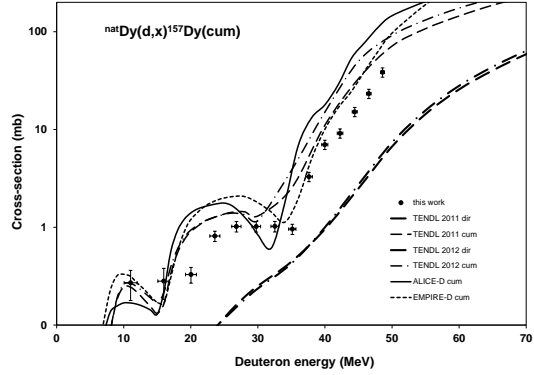}
\caption{Experimental and theoretical excitation functions for $^{nat}$Dy(d,x)$^{157}$Dy(cum)}
\end{figure}

\subsubsection{Cross-sections for the $^{nat}$Dy(d,x)$^{155}$Dy (cum) reaction}
\label{4.1.6}
The cumulative cross-section of the $^{155}$Dy (T$_{1/2}$ = 9.9 h) represents the sum of the direct and indirect production through the decay of 155Ho (T$_{1/2}$ = 48 min). The agreement with the results of the three codes is acceptable. (Fig. 7)

\begin{figure}[h]
\includegraphics[scale=0.3]{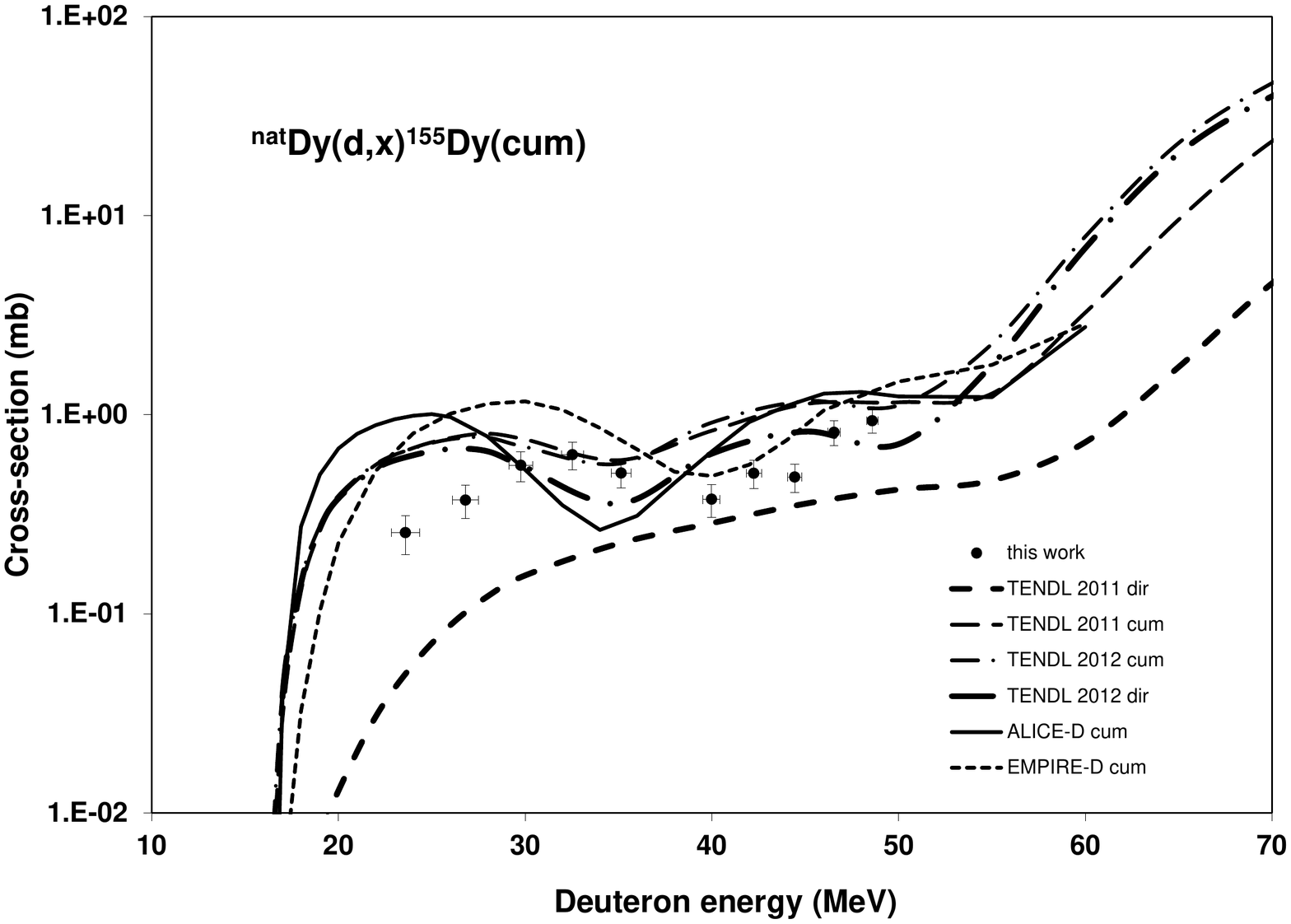}
\caption{Experimental and theoretical excitation functions for $^{nat}$Dy(d,x)$^{155}$Dy(cum)}
\end{figure}

\subsubsection{Cross-sections for the $^{nat}$Dy(d,x)$^{161}$Tb reaction}
\label{4.1.7}
The $^{161}$Tb (T$_{1/2}$ = 6.89 d) is produced via direct (d,2pxn) reactions if we neglect the small contribution from 161Gd decay (T$_{1/2}$ = 3.66 min) produced with the (d,3pxn) reaction. Our experimental data above 30 MeV are systematically higher than the theoretical results and the bump predicted by EMPIRE is not reproduced by the experiment (Fig. 8).

\begin{figure}[h]
\includegraphics[scale=0.3]{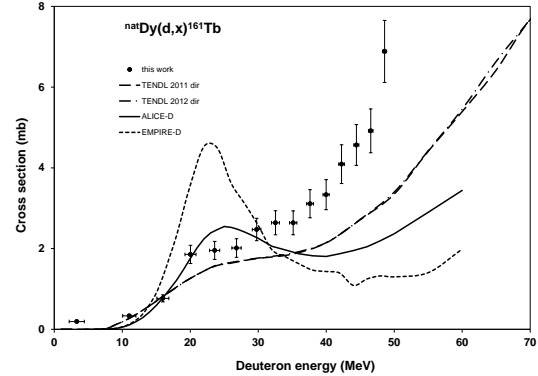}
\caption{Experimental and theoretical excitation functions for $^{nat}$Dy(d, x)$^{161}$Tb}
\end{figure}

\subsubsection{Cross-sections for the $^{nat}$Dy(d,x)$^{160}$Tb reaction}
\label{4.1.8}
The radionuclide $^{160}$Tb (T$_{1/2}$ = 72.3 d) can only be produced directly by (d,2pxn) reactions as decay contributions are closed from both sides. The activation cross-sections are shown in Fig. 9.  There are significant disagreements with the results of the theoretical codes.

\begin{figure}[h]
\includegraphics[scale=0.3]{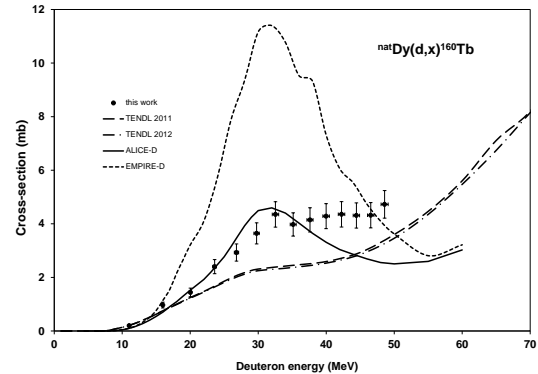}
\caption{Experimental and theoretical excitation functions for $^{nat}$Dy(d,x) $^{160}$Tb}
\end{figure}

\subsubsection{Cross-sections for the $^{nat}$Dy(d,x)$^{156}$Tb(cum)  reaction}
\label{4.1.9}
The production cross-sections of $^{156}$Tb (Fig. 10) represent total activation cross-sections of the ground state:  i.e. cumulative cross-section of the ground state after the decay of the two isomeric states ($^{156m1}$Tb, IT: 100 \%, T$_{1/2}$ = 3 h and $^{156m2}$Tb, IT: 100 \%, T$_{1/2}$ = 24.4 h). Large divergences can be seen between the different codes and the experimental results.

\begin{figure}[h]
\includegraphics[scale=0.3]{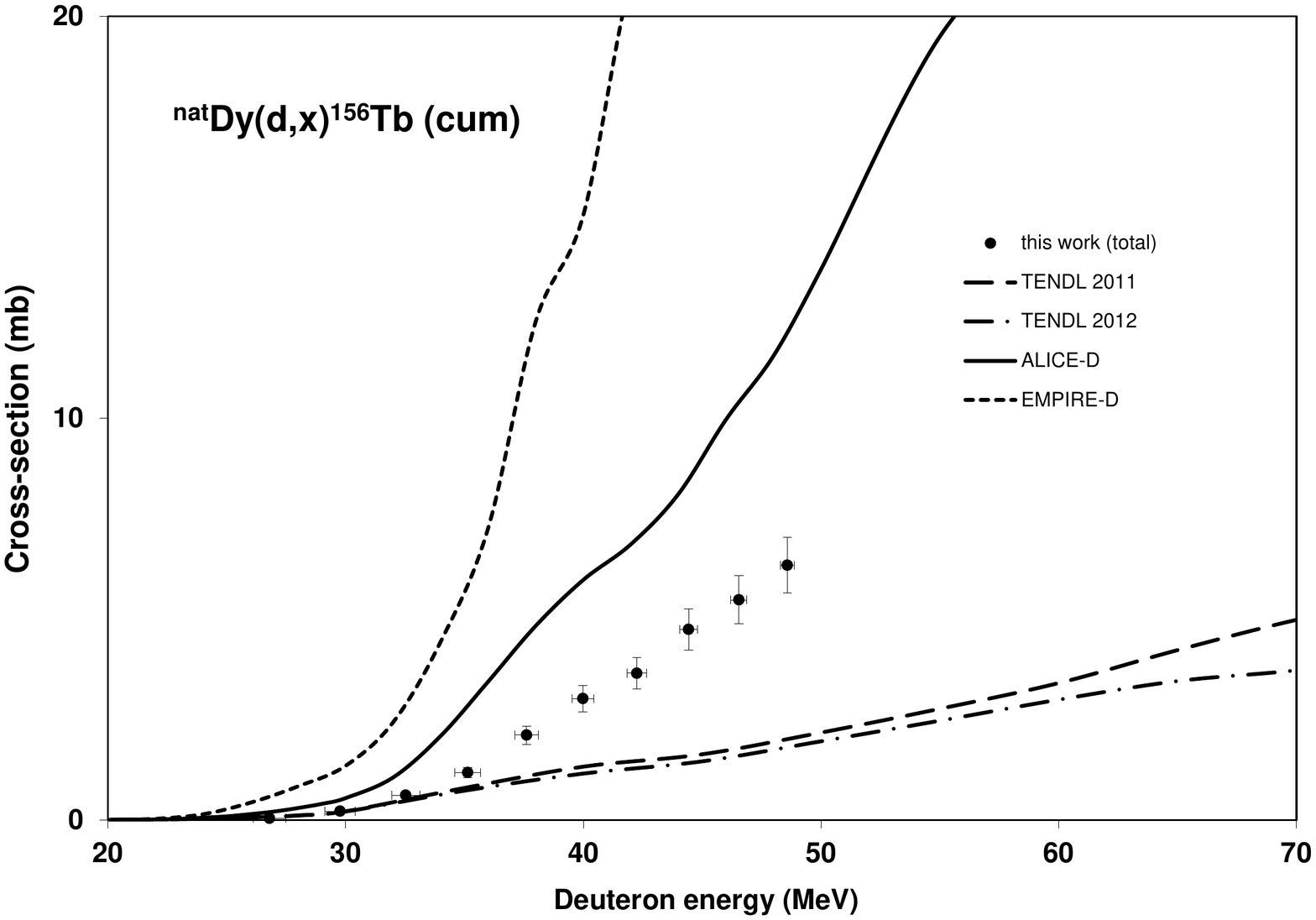}
\caption{Experimental and theoretical excitation functions for $^{nat}$Dy(d,x) $^{156}$Tb(cum)}
\end{figure}

\subsubsection{Cross-sections for the $^{nat}$Dy(d,x)$^{155}$Tb(cum) reaction}
\label{4.1.10}
The cross-sections for $^{155}$Tb (T$_{1/2}$ = 5.32 d) production are cumulative. They were determined from the spectra measured after the decay of progenitor $^{155}$Ho (T$_{1/2}$ = 48 min) and $^{155}$Dy (T$_{1/2}$ = 9.9 h) isotopes. The best description is given by the TENDL calculations, while up to 35 MeV the ALICE-D is also acceptable (Fig. 11).

\begin{figure}[h]
\includegraphics[scale=0.3]{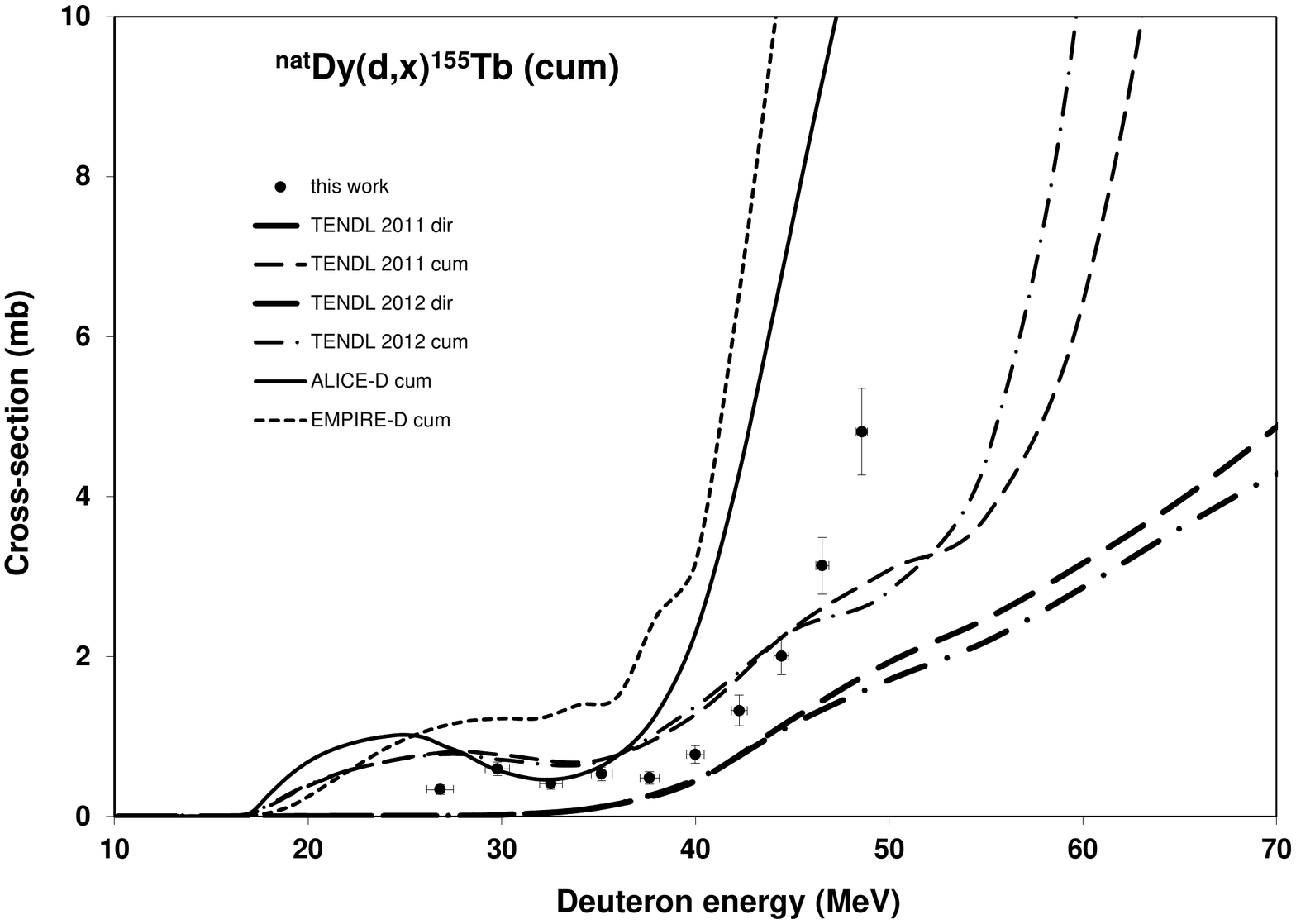}
\caption{Experimental and theoretical excitation functions for $^{nat}$Dy(d, x) $^{155}$Tb(cum)}
\end{figure}

\subsection{Integral yields}
\label{4.2}
We calculated the integral yields for all investigated radionuclides from a spline fit to our experimental data. The yields represent so called physical yields for instantaneous irradiation. No data for directly measured thick target yields were found in the literature. The yields are presented in Figs. 12 and 13. In Fig. 13 the most interesting isotopes are the $^{161}$Ho and $^{162m}$Ho, as they have also medical interest (Tárkányi et al., 2013). The $^{165g}$Dy production yield is with an order of magnitude lower than that of the above two. Another group is formed from the remaining dysprosium isotopes i.e. $^{159,157,155}$Dy, which have much lower yield. In Fig. 13 the terbium radio-isotopes are presented. From the point of view of yield they rather belong to the latter group of the preceding figure. 

\begin{figure}[h]
\includegraphics[scale=0.3]{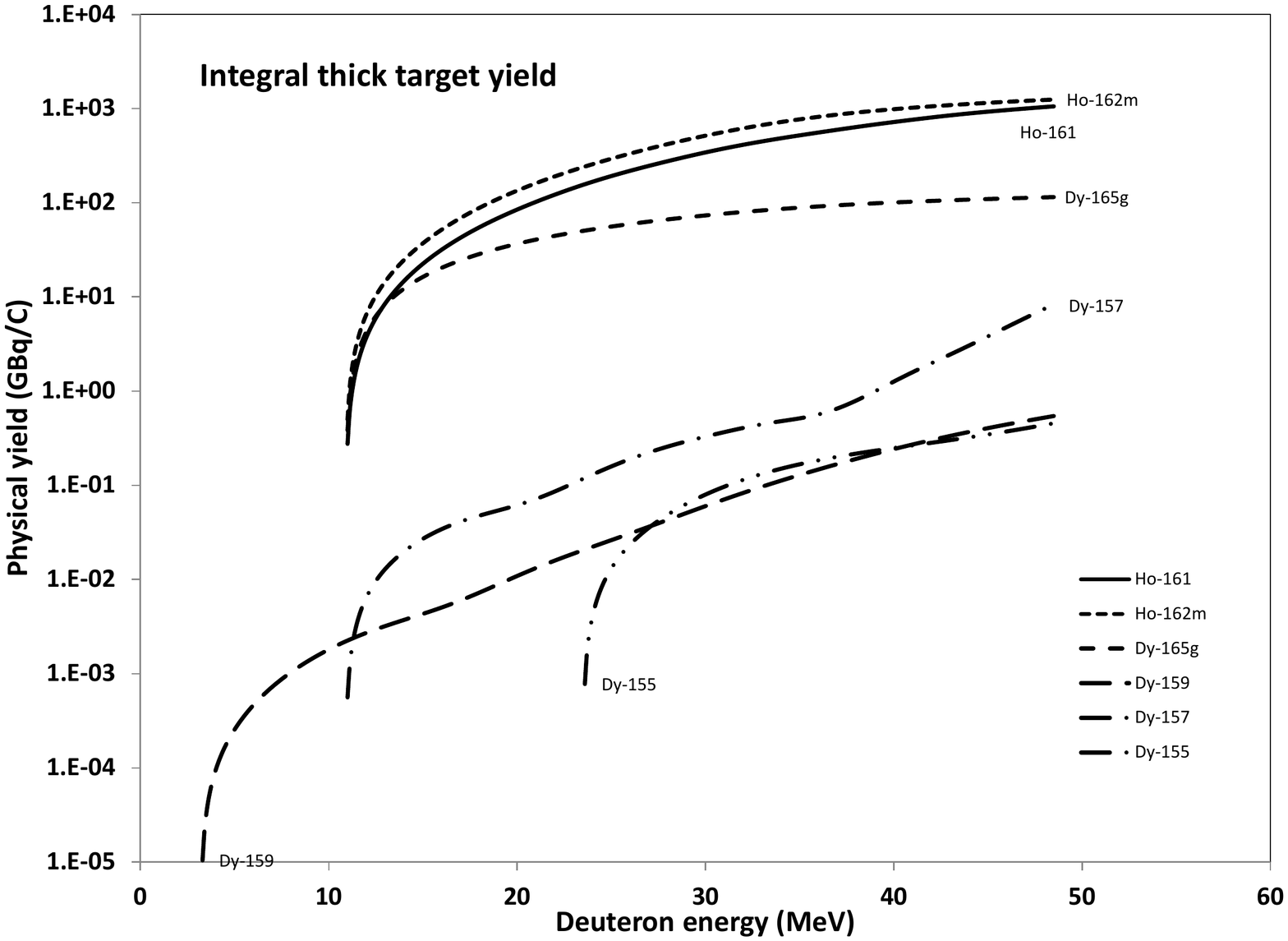}
\caption{Integral yields as a function of proton energy for the production of $^{162m,161}$Ho, $^{165,159,157,155}$Dy radioisotopes }
\end{figure}

\begin{figure}[h]
\includegraphics[scale=0.3]{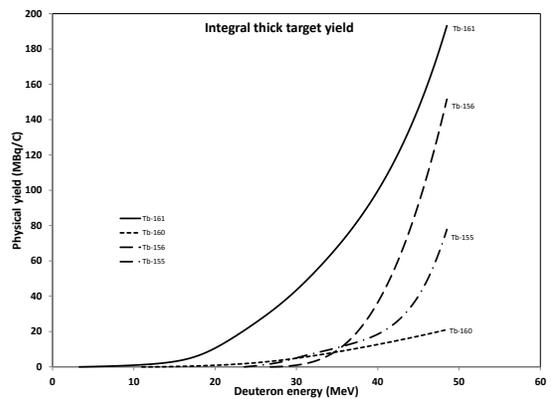}
\caption{Integral yields as a function of proton energy for the production of  $^{161,160,156,155}$Tb radioisotopes.}
\end{figure}

\begin{table*}[t]
\tiny
\caption{Experimental cross-sections of ${}^{nat}$Dy(d,x)${}^{162m}$Ho, ${}^{161}$${}^{g}$Ho(m+), ${}^{165g}$Dy(m+), ${}^{159}$Dy(cum),${}^{ 157}$Dy(cum),${}^{ 155}$Dy(cum)  reactions}
\centering
\begin{center}
\begin{tabular}{|p{0.3in}|p{0.3in}|p{0.3in}|p{0.3in}|p{0.3in}|p{0.3in}|p{0.3in}|p{0.3in}|p{0.3in}|p{0.3in}|p{0.3in}|p{0.3in}|p{0.3in}|p{0.3in}|} \hline 
\multicolumn{2}{|p{1in}|}{E $\pm\Delta$E\newline (MeV)} & \multicolumn{12}{|p{3.8in}|}{Cross-section($\sigma$)$\pm\Delta\sigma$\newline (mb)} \\ \hline 
\multicolumn{2}{|p{1in}|}{} & \multicolumn{2}{|p{0.7in}|}{${}^{162m}$Ho} & \multicolumn{2}{|p{0.5in}|}{${}^{161g}$Ho} & \multicolumn{2}{|p{0.7in}|}{${}^{165g}$Dy} & \multicolumn{2}{|p{0.7in}|}{${}^{159}$Dy} & \multicolumn{2}{|p{0.7in}|}{${}^{157}$Dy} & \multicolumn{2}{|p{0.6in}|}{${}^{155}$Dy} \\ \hline 
3.3 & 1.1 &  & ~ &  & ~ &  & ~ & 4.8 & 0.5 &  & ~ &  & ~ \\ \hline 
11.0 & 1.0 & 36.6 & 4.2 & 45.0 & 5.5 & 58.0 & 6.4 & 10.4 & 1.1 & 0.3 & 0.1 &  & ~ \\ \hline 
16.0 & 0.9 & 91.7 & 10.1 & 127.5 & 14.7 & 56.6 & 6.2 & 14.4 & 1.6 & 0.3 & 0.1 &  & ~ \\ \hline 
20.0 & 0.8 & 126.1 & 13.9 & 181.0 & 20.5 & 40.4 & 4.6 & 34.2 & 3.7 & 0.3 & 0.1 &  & ~ \\ \hline 
23.6 & 0.7 & 147.5 & 16.3 & 234.9 & 26.3 & 32.8 & 3.8 & 45.9 & 5.0 & 0.8 & 0.1 & 0.3 & 0.1 \\ \hline 
26.8 & 0.7 & 172.8 & 19.0 & 250.9 & 28.3 & 27.5 & 3.4 & 70.6 & 7.6 & 1.0 & 0.1 & 0.4 & 0.1 \\ \hline 
29.8 & 0.6 & 189.9 & 20.8 & 295.1 & 33.1 & 26.9 & 3.3 & 118.9 & 12.9 & 1.0 & 0.1 & 0.6 & 0.1 \\ \hline 
32.5 & 0.6 & 192.5 & 21.1 & 277.7 & 31.3 & 25.9 & 3.1 & 156.7 & 17.0 & 1.0 & 0.1 & 0.6 & 0.1 \\ \hline 
35.1 & 0.5 & 166.8 & 18.4 & 265.8 & 29.8 & 20.1 & 2.7 & 165.4 & 17.9 & 1.0 & 0.1 & 0.5 & 0.1 \\ \hline 
37.6 & 0.5 & 143.2 & 15.7 & 301.0 & 33.7 & 18.6 & 2.4 & 237.5 & 25.7 & 3.3 & 0.4 &  &  \\ \hline 
40.0 & 0.5 & 106.1 & 12.5 & 274.8 & 31.3 & 10.5 & 1.9 & 290.3 & 31.4 & 7.0 & 0.8 & 0.4 & 0.1 \\ \hline 
42.3 & 0.4 & 93.9 & 10.8 & 295.2 & 33.1 & 10.3 & 1.7 & 298.8 & 32.3 & 9.1 & 1.0 & 0.5 & 0.1 \\ \hline 
44.4 & 0.4 & 94.4 & 11.2 & 242.0 & 28.1 & 9.6 & 2.0 & 315.3 & 34.1 & 15.2 & 1.6 & 0.5 & 0.1 \\ \hline 
46.5 & 0.3 & 75.2 & 10.0 & 222.2 & 26.9 & 8.8 & 3.0 & 320.3 & 34.7 & 23.2 & 2.5 & 0.8 & 0.1 \\ \hline 
48.6 & 0.3 & 66.7 & 9.0 & 219.9 & 26.6 & 10.0 & 2.6 & 356.0 & 38.5 & 38.5 & 4.2 & 0.9 & 0.1 \\ \hline 
\end{tabular}
\end{center}
\end{table*}

\begin{table*}[t]
\tiny
\caption{Experimental cross-sections of ${}^{nat}$Dy(d,x)${}^{161}$Tb, ${}^{160}$Tb, ${}^{156}$Tb(cum),${}^{ 155}$Tb(cum) reactions}
\centering
\begin{center}
\begin{tabular}{|p{0.3in}|p{0.3in}|p{0.3in}|p{0.3in}|p{0.3in}|p{0.3in}|p{0.3in}|p{0.3in}|p{0.3in}|p{0.3in}|} \hline 
\multicolumn{2}{|p{1in}|}{E $\pm\Delta$E\newline (MeV)} & \multicolumn{8}{|p{2.5in}|}{Cross-section($\sigma$)$\pm\Delta\sigma$\newline (mb)} \\ \hline 
\multicolumn{2}{|p{1in}|}{} & \multicolumn{2}{|p{0.7in}|}{${}^{161}$Tb} & \multicolumn{2}{|p{0.5in}|}{${}^{160}$Tb} & \multicolumn{2}{|p{0.7in}|}{${}^{156}$Tb} & \multicolumn{2}{|p{0.7in}|}{${}^{155}$Tb} \\ \hline 
3.3 & 1.1 & 0.19 & 0.02 &  & ~ &  & ~ &  &  \\ \hline 
11.0 & 1.0 & 0.3 & 0.01 & 0.20 & 0.02 &  & ~ &  &  \\ \hline 
16.0 & 0.9 & 0.8 & 0.1 & 1.0 & 0.1 &  & ~ &  &  \\ \hline 
20.0 & 0.8 & 1.9 & 0.2 & 1.4 & 0.2 &  & ~ &  &  \\ \hline 
23.6 & 0.7 & 2.0 & 0.2 & 2.4 & 0.3 &  & ~ & 0.2 & 0.01 \\ \hline 
26.8 & 0.7 & 2.0 & 0.2 & 2.9 & 0.3 & 0.04 & 0.01 & 0.3 & 0.1 \\ \hline 
29.8 & 0.6 & 2.5 & 0.3 & 3.6 & 0.4 & 0.22 & 0.03 & 0.6 & 0.1 \\ \hline 
32.5 & 0.6 & 2.6 & 0.3 & 4.4 & 0.5 & 0.6 & 0.1 & 0.4 & 0.1 \\ \hline 
35.1 & 0.5 & 2.6 & 0.3 & 4.0 & 0.4 & 1.2 & 0.1 & 0.5 & 0.1 \\ \hline 
37.6 & 0.5 & 3.1 & 0.3 & 4.1 & 0.5 & 2.1 & 0.2 & 0.5 & 0.1 \\ \hline 
40.0 & 0.5 & 3.3 & 0.4 & 4.3 & 0.5 & 3.0 & 0.3 & 0.8 & 0.1 \\ \hline 
42.3 & 0.4 & 4.1 & 0.5 & 4.4 & 0.5 & 3.7 & 0.4 & 1.3 & 0.2 \\ \hline 
44.4 & 0.4 & 4.6 & 0.5 & 4.3 & 0.5 & 4.7 & 0.5 & 2.0 & 0.2 \\ \hline 
46.5 & 0.3 & 4.9 & 0.5 & 4.3 & 0.5 & 5.5 & 0.6 & 3.1 & 0.4 \\ \hline 
48.6 & 0.3 & 6.9 & 0.8 & 4.7 & 0.5 & 6.3 & 0.7 & 4.8 & 0.5 \\ \hline 
\end{tabular}
\end{center}
\end{table*}

\section{Summary and conclusions}
\label{5}
In the frame of systematic investigation of excitation functions of deuteron induced nuclear reactions, activation cross-sections on dysprosium were investigated up to 50 MeV. Independent or cumulative cross-sections for the formation of the radionuclides $^{162m,161}$Ho,$^{165,159,157,155}$Dy and $^{161,160,156,155}$Tb through $^{nat}$Dy(d,x) nuclear reactions were measured for the first time.
The experimental data were compared with the theoretical data obtained by the EMPIRE-D and the ALICE-IPPE-D codes and the TALYS data in TENDL library. The prediction of the codes differs significantly from each other, and from the experimental data. In the case of EMPIRE-D and ALICE-D the phenomenological systematics improved but not solved all the disagreements. In the case of deuteron induced reactions there are still no significant improvements between the different TENDL versions. The (d,p) results of the  TALYS code in TENDL are still largely discrepant. The new experimental data are therefore very important, to complete experimental activation database of deuteron induced reactions and for further improvement of theoretical models. 
As it was mentioned, the dysprosium is an important technological material, therefore the  measured activation cross-sections can be useful in the field of medical isotope production and to estimate the radiation dose of  caused by primer or secondary deuterons in  reactors, accelerators and in space applications.

\section{Acknowledgements}
\label{7}
This work was performed in the frame of the HAS-FWO Vlaanderen (Hungary-Belgium) project. The authors acknowledge the support of the research project and of the respective institutions. We are grateful to all the authorities concerned. We thank to the Cyclotron Laboratory of the Université Catholique in Louvain la Neuve (LLN) providing the beam time and the staff of the LLN Cyclone 90 cyclotron for performing the irradiations
 


%
%

\clearpage
\bibliographystyle{elsarticle-harv}
\bibliography{Dyd}




%



\end{document}